\documentclass{moriond}

\def\be{\begin{equation}}
\def\ee{\end{equation}}
\def\bea{\begin{eqnarray}}
\def\eea{\end{eqnarray}}

\newcommand{\Photo}{}

\begin{document}
\vspace*{4cm}
\title{A Standard Model analysis of \boldmath $D^0 \to \pi^- \pi^+, K^- K^+, K_{\rm S}^0 K_{\rm S}^0$ }

\author{Robert Fleischer$^{a,b}$, \underline{Maria Laura Piscopo}$^{c}$, K. Keri Vos$^{a,d}$, B. Ya\u{g}mur Zubaro\u{g}lu$^{a}$}
\author{ }

\address{${}^a$  Nikhef, Science Park 105, NL-1098 XG Amsterdam,  Netherlands\\
${}^b$  Department of Physics and Astronomy, Vrije Universiteit Amsterdam,\\
NL-1081 HV Amsterdam, Netherlands\\
${}^c$Theoretical Physics Department, CERN,
1211 Geneva 23, Switzerland\\
$^d$Gravitational Waves and Fundamental Physics (GWFP),\\ 
Maastricht University, Duboisdomein 30,\\
NL-6229 GT Maastricht, Netherlands}

\maketitle\abstracts{
We investigate the Standard Model~(SM) description of the singly Cabibbo-suppressed charm decays $D^0\to \pi^-\pi^+$, $D^0\to K^-K^+$ and $D^0\to K_{\rm S}^0K_{\rm S}^0$. Using factorisation together with isospin symmetry, we constrain non-factorisable effects and the associated $U$-spin breaking required by the measured branching fractions. We find that corrections of order $50\%$ are sufficient to describe all modes, which although sizeable, are not unexpected for hadronic charm decays. Using the constraints from the branching fractions, we derive SM predictions for the direct CP asymmetry in $D^0\to K_{\rm S}^0K_{\rm S}^0$, finding it to be at most at the per-mille level in our benchmark scenario, providing motivation for future precision measurements.}

\section{Introduction}
Charm decays provide a unique probe of the interplay between weak and strong interactions in the SM. However, obtaining robust theoretical predictions remains challenging~\cite{Friday:2025gpj}.

Among hadronic charm decays, the singly Cabibbo-suppressed modes $D^0\to\pi^-\pi^+$, $D^0\to K^-K^+$ and $D^0\to K_{\rm S}^0K_{\rm S}^0$ are of particular interest. Although they proceed through the same underlying quark-level transitions, they probe different hadronic dynamics and $U$-spin-breaking effects. The $D^0\to\pi^-\pi^+$ and $D^0\to K^-K^+$ modes are related by $U$-spin symmetry and admit a factorisation description, as they receive contributions from colour-allowed tree amplitudes. Their branching fractions are precisely measured and exhibit sizeable $U$-spin breaking~\cite{ParticleDataGroup:2026aaa}. By contrast, $D^0\to K_{\rm S}^0K_{\rm S}^0$ proceeds only via non-factorisable contributions, dominated by the exchange topology, and essentially vanishes in the $U$-spin limit. Nevertheless, its branching fraction is only about an order of magnitude smaller~\cite{ParticleDataGroup:2026aaa} than those of $D^0\to\pi^-\pi^+$ and $D^0\to K^-K^+$. These complementary dynamics are also reflected in the pattern of direct CP violation.

Experimentally, CP violation in charm has been established at the per-mille level by the LHCb collaboration through the measurement of the difference of CP asymmetries, $\Delta A_{\rm CP}$, between $D^0\to K^-K^+$ and $D^0\to\pi^-\pi^+$~\cite{LHCb:2019hro}. More recently, LHCb reported the first evidence for CP violation in $D^0\to\pi^-\pi^+$~\cite{LHCb:2022lry}. On the other hand, CP violation in $D^0\to K_{\rm S}^0K_{\rm S}^0$ remains less constrained, with current measurements consistent with zero at the percent level~\cite{Punzi}.

In this contribution, we discuss how the measured branching fractions can be used to constrain non-factorisable contributions and associated $U$-spin breaking effects in the SM, and to derive bounds on the direct CP asymmetry in $D^0\to K_{\rm S}^0K_{\rm S}^0$. Our analysis is based on the factorisation framework and supplemented by minimal additional assumptions~\cite{Fleischer:2025zhl}.

\section{\boldmath The decays $D^0 \to K^- K^+$ and $D^0 \to \pi^- \pi^+$}
\begin{figure}
\centering
\begin{minipage}{0.58\textwidth}
\includegraphics[scale = 0.48]{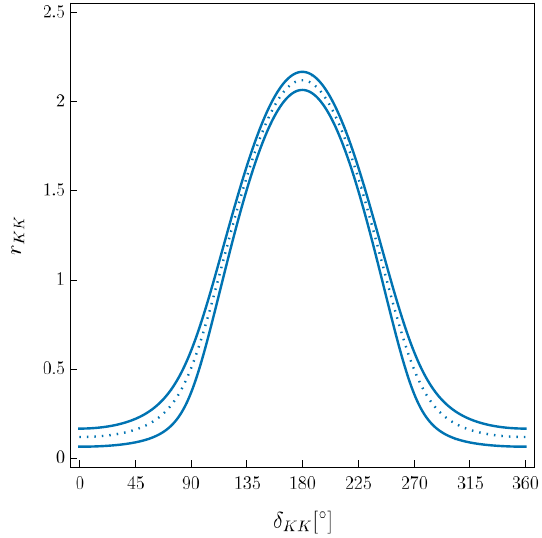}
\includegraphics[scale = 0.48]{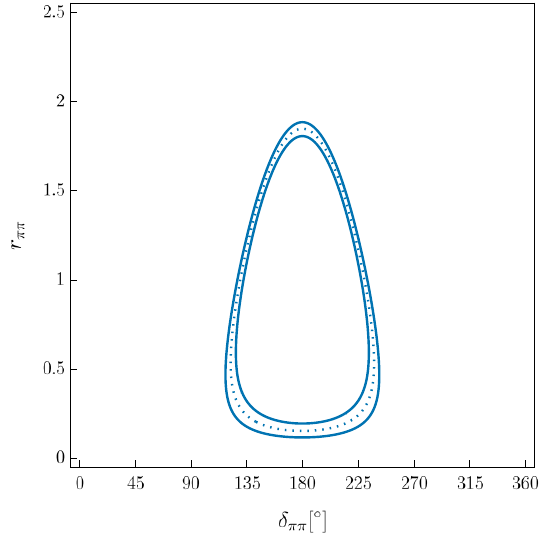}
\caption{Allowed $1\sigma$  ranges for the parameters $r_f$ and $\delta_f$ defined in Eq.~\ref{eq:Ampl-decomposition} for $f = \pi\pi$~(left) and $f = KK$ (right).}
\label{fig:2}
\end{minipage}
\quad
\begin{minipage}{0.36\textwidth}
\includegraphics[scale = 0.63]{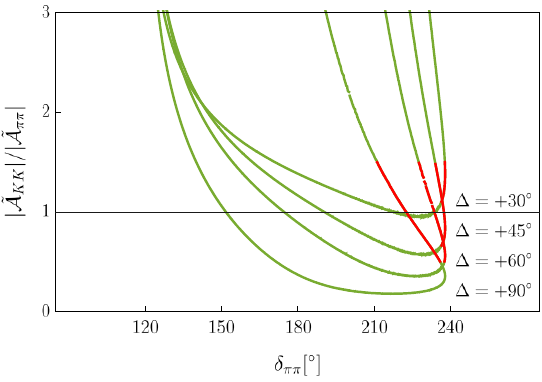}
\caption{Allowed values of $|\tilde {\cal A}_{KK}/\tilde {\cal A}_{\pi\pi}|$ as a function of $\delta_{\pi\pi}$. The red region corresponds to the constraints in Eq.~\ref{eq:constraints_P1P2}.}
\label{fig:2b}
\end{minipage}
\end{figure}

Within the SM, the decays $D^0\to\pi^-\pi^+$ and $D^0\to K^-K^+$ receive contributions from colour-allowed tree ($T$), exchange ($E$), and penguin ($P_q$) topologies. For $D^0\to K^-K^+$, the unitarity of the Cabibbo-Kobayashi-Maskawa~(CKM) matrix, $\lambda_d+\lambda_s+\lambda_b=0$, where $\lambda_q\equiv V_{cq}^*V_{uq}$ ($q=d,s,b$), allows the decay amplitude to be written as
\begin{equation}
A(D^0 \to K^-K^+) = \lambda_s \;{\cal A}_{KK} \left(1+ \frac{\lambda_b}{\lambda_s}  \frac{P_{bd}}{ T + E + P_{sd}}\right)\,,
\quad \mbox{with} \quad 
  \mathcal{A}_{KK} \equiv T + E + P_{sd} \, , 
\label{eq:AK2}
\end{equation}
and $P_{qr} \equiv P_q - P_r$. The corresponding expressions for $D^0\to\pi^-\pi^+$ follow from $T \to T^\prime, E \to E^\prime, P_{sd} \to - P_{sd}^\prime$ and $\lambda_s \to \lambda_d$, where with the primes we indicate the $c \to d$ transition.
Such parametrisations have long been used in the $B$ sector~\cite{Fleischer:1999pa}. However, in charm decays the CKM hierarchy is stronger: $\lambda_s \sim -\lambda_d \sim \lambda$, while $|\lambda_b/\lambda_{d,s}|\sim\lambda^4 $, where $\lambda \sim 0.22$ is the Wolfenstein parameter~\cite{Charles:2004jd}. Consequently, CP-averaged branching fractions are dominated by the leading CKM amplitudes ${\cal A}_{KK}$ and ${\cal A}_{\pi \pi}$. 
The latter can be estimated in naive factorisation, where
\begin{equation} 
T \to T^{\rm fac}_{KK}\equiv i \frac{G_F}{\sqrt{2}}  a_1 f_{K} \left( m_{D^0}^2 - m_{K}^2 \right) f_0^{D \to K}(m_{K}^2)\,.
\label{eq:T_fac}
\end{equation}
Here $a_1 \sim 1.1$ is a linear combination of current-current operators Wilson coefficients and $f_0$ and $f_K$ denote the scalar $D \to K$ form factor and the kaon decay constant. An analogous expression holds for $T^{\rm fac}_{ \pi \pi}$. 
Using precise lattice QCD determinations of decay constants~\cite{FlavourLatticeAveragingGroupFLAG:2024oxs} and form factors~\cite{FermilabLattice:2022gku}, we obtain for the branching fractions in the factorisation limit~\cite{Fleischer:2025zhl}:
\begin{equation} 
{\cal B}(D^0 \to K^- K^+)|_{\rm fac}    = (3.27^{+0.36}_{-0.27}) \times 10^{-3}\,,
\quad 
{\cal B}(D^0 \to \pi^-\pi^+)|_{\rm fac}   \,\,\, = (2.03^{+0.23}_{-0.17}) \times 10^{-3}\,,
\label{eq:Br_fac_FNAL}
\end{equation}
to be compared with the experimental measurements~\cite{ParticleDataGroup:2026aaa}
\begin{equation} 
{\cal B}(D^0 \to K^- K^+)|_{\rm exp}    = (4.08 \pm 0.06) \times 10^{-3}\,,
\,\,\,\,
{\cal B}(D^0 \to \pi^-\pi^+)|_{\rm exp}   \,\,\, = (1.454 \pm 0.024) \times 10^{-3}\,.
\end{equation}
The factorisation predictions are of the correct order of magnitude and in reasonable agreement with the experimental values, as previously noted~\cite{Lenz:2023rlq}, suggesting non-factorisable corrections at the (20-30)$\%$ level, with opposite signs in the two channels. 
To quantify these effects, we write
\begin{equation} 
{\cal A}_{f} =  T_{f}^{\rm fac} + \tilde {\cal A}_f  = T_{f}^{\rm fac}  \left( 1 + r_{f} e^{i \delta_{f}} \right)\,, \qquad 
r_f e^{i\delta_f}\equiv \frac{ \tilde {\cal A}_{f} }{ T_{f}^{\rm fac}}\,,  \qquad \mbox{with} \quad f = KK, \pi \pi\,,
\label{eq:Ampl-decomposition}
\end{equation}
where $r_f$ denotes the magnitude of the non-factorisable contribution relative to the factorisation result. Neglecting subleading CKM corrections of the order $|\lambda_b/\lambda_{d,s}| \sim10^{-3}$, we obtain
\begin{equation}
{\cal B}(D^0 \to f) = \Big[ 1 + r_{f}^2 + 2 r_{f} \cos \delta_{f} \Big] {\cal B}(D^0 \to f)|_{\rm fac} \,.
\label{eq:Br-P1mP2p}
\end{equation}
A fit to the measured branching fractions yields the allowed regions in the $(\delta_f,r_f)$ plane shown in Fig.~\ref{fig:2}~\cite{Fleischer:2025zhl}. Although the system is under constrained, the data are well described by moderate non-factorisable effects, with $r_f\sim0.5$. Moreover, it is possible to identify scenarios with simultaneously moderate non-factorisable contributions and moderate non-factorisable $U$-spin-breaking effects, where the latter is quantified by $1 -\left| \tilde {\cal A}_{KK}/\tilde {\cal A}_{\pi\pi} \right|$ ~\cite{Fleischer:2025zhl}:
\begin{equation}
    r_{KK} \leq 0.7\,, \quad r_{\pi\pi} \leq 0.7\,, \quad
   0.5 \leq  \left| \frac{\tilde {\cal A}_{KK}}{\tilde {\cal A}_{\pi\pi}} \right| \leq 1.5\,.
    \label{eq:constraints_P1P2}
\end{equation}
This is shown in Fig.~\ref{fig:2b}, for fixed $U$-spin breaking in the relative strong phases $\Delta \equiv \delta_{KK} -\delta_{\pi\pi} $. 
\section{The decay \boldmath $D^0 \to K_{\rm S}^0 K_{\rm S}^0$}
\begin{figure}
\centering
\begin{minipage}{0.485\textwidth}
    \centering
    \includegraphics[width=0.7\textwidth]{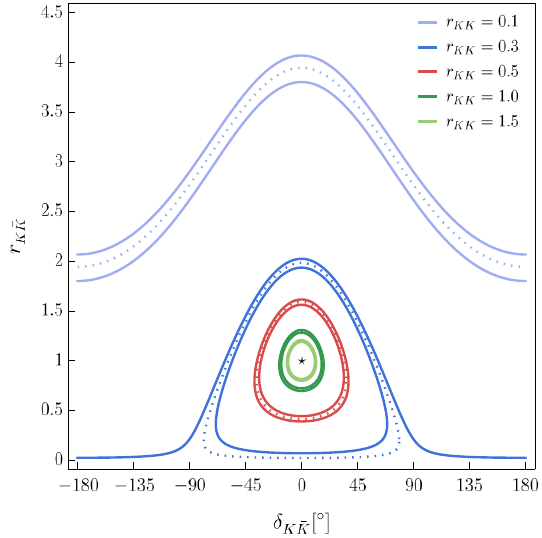}
    \caption{Allowed $1\sigma$ ranges for the $U$-spin-breaking parameters $r_{K\bar K}$ and $\delta_{K \bar K}$.}
    \label{fig:3a}
\end{minipage}
\hfill
\begin{minipage}{0.48\textwidth}
    \centering
    \includegraphics[width=0.95\textwidth]{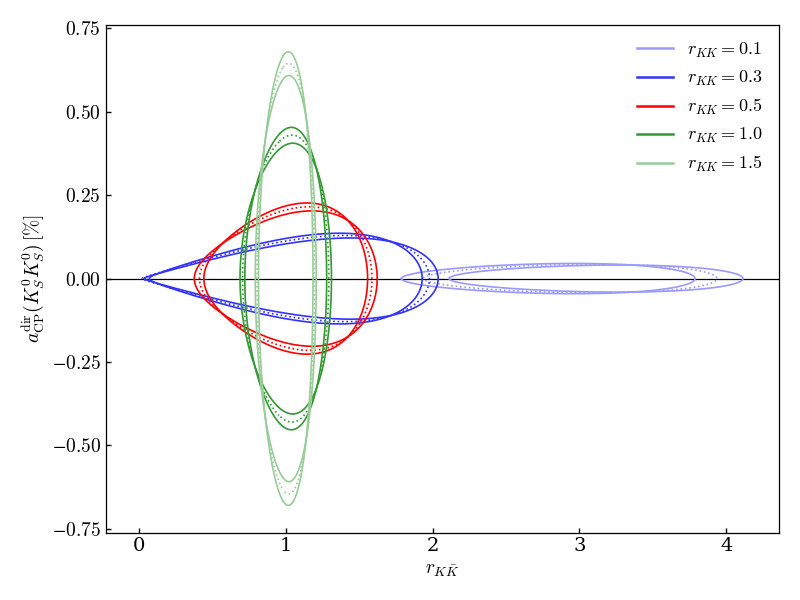}
    \caption{Direct CP asymmetry in $D^0 \to K_{\rm S}^0 K_{\rm S}^0$ as a function of $r_{K\bar{K}}$ for fixed values of $r_{KK}$.}
    \label{fig:3b}
\end{minipage}
\end{figure}

The decay $D^0\to K^0\bar K^0$~\footnote{We define $|K_{\rm S}^0 \rangle \equiv (|K^0\rangle - |\bar K^0\rangle)/\sqrt{2}$, which neglects tiny effects from CP violation in the kaon sector at the level of $10^{-3}$. The branching ratio for $D^0 \to K_{\rm S}^0 K^0_{\rm S}$ is obtained as
${\cal B}(D^0 \to K^0_{\rm S} K^0_{\rm S}) = \frac12 {\cal B}(D^0 \to K^0 \bar K^0)\,.$} has a qualitatively different structure: tree and penguin topologies are absent, leaving only exchange and penguin-annihilation~($PA_q$) contributions,
\begin{equation}
\label{eq:AK0K0formula}
     A(D^0\to K^0\bar{K}^0) = \lambda_s \; \mathcal{A}_{K\bar{K}} \left[1+ \frac{\lambda_b}{\lambda_s} a_{K\bar{K}} e^{i\theta_{K\bar{K}}}  \right] \,,
\end{equation}
where
\begin{equation}
    \mathcal{A}_{K\bar{K}} = (E_d - E_s) + PA_{sd} \, ,   \qquad a_{K\bar{K}} \; e^{i\theta_{K\bar{K}}} \equiv \frac{- E_s + PA_{bd}}{(E_d - E_s) + PA_{sd}}\,.
\end{equation}
In the $U$-spin limit, $E_s=E_d$ and $PA_{sd} \equiv PA_s - PA_d=0$, such that the decay is strongly suppressed by $|\lambda_b/\lambda_s|^2$. Moreover, being purely non-factorisable, the decay amplitude vanishes in the factorisation approximation. Nevertheless, experimentally~\cite{ParticleDataGroup:2026aaa}:
\begin{equation} 
{\cal B}(D^0 \to K_{\rm S}^0 K_{\rm S}^0)|_{\rm exp} = (1.41 \pm 0.05) \times 10^{-4}\,.
\label{eq:Br_KsKs_exp-0}
\end{equation}
To investigate the implications of this measurement, we write
\begin{equation}
{\cal A}_{K\bar{K}}  =  (E_d + PA_{s}) \left( 1 - r_{K \bar K} e^{i \delta_{K \bar K}} \right)\,, \quad 
r_{K \bar K} \; e^{i \delta_{K\bar{K}}} \equiv \frac{ E_s + PA_d}{E_d + PA_s}\,,
\label{eq:rKKbar_deltaKKbar}
\end{equation}
where $r_{K\bar K}=1$ and $\delta_{K\bar K}=0$ in the exact $U$-spin limit. To proceed, we make two assumptions that simply the expressions and allow us to obtain constraints on the size of non-factorisable contributions and $U$-spin breaking. 
First, we neglect penguin-annihilation topologies, expected to be loop- and colur-suppressed. Second, we assume that the non-factorisable contributions in $D^0\to K^-K^+$, parametrised by $r_{KK}$, are dominated by the exchange topology. Under these assumptions, $|E_d|$ in Eq.~\ref{eq:rKKbar_deltaKKbar} can be inferred from $r_{KK}$, since the corresponding exchange amplitudes are related by $d\leftrightarrow u$ and are equal in the isospin limit. This leads to
\begin{equation}
{\cal B}(D^0 \to K^0_{\rm S} K^0_{\rm S}) = \frac{1}{2} \Big[ 1 + r_{K \bar K}^2 - 2 r_{K \bar K} \cos \delta_{K \bar K} \Big] \,  {\cal B}(D^0 \to K^- K^+)|_{\rm fac} \, r_{K K}^2 \,,
\label{eq:Br_KsKs}
\end{equation}
where the factorised branching fraction is taken from Eq.~\ref{eq:Br_fac_FNAL}.
The allowed regions in the $(\delta_{K\bar K}, r_{K\bar K})$ plane obtained from the measured branching fraction are shown in Fig.~\ref{fig:3a} for representative values $r_{KK} = \{0.1, 0.3, 0.5, 1.0, 1.5\}$. Larger values of $r_{KK}$ require smaller $U$-spin breaking effects, whereas $r_{KK}\sim0.1$ would imply effects of order $100\%$. The scenario $r_{KK}\sim0.5$, favoured by the analysis of $D^0\to K^-K^+$ and $D^0\to\pi^-\pi^+$, corresponds to moderate breaking of order $50\%$~\cite{Fleischer:2025zhl}. 

Finally, the direct CP asymmetry in $D^0\to K_{\rm S}^0K_{\rm S}^0$ is given by:
\begin{equation}
a_{\rm CP}^{\rm dir} (K^0_{\rm S} K^0_{\rm S}) =   2 \left| \frac{\lambda_b}{\lambda_s} \right| \sin \gamma \left| \frac{r_{K \bar K} e^{i \delta_{K \bar K}}  }{1- r_{K \bar K} e^{i \delta_{K \bar K}} } \right|  \sin \left[ \arg \left( \frac{ r_{K \bar K} e^{i \delta_{K \bar K}}  }{1- r_{K \bar K} e^{i \delta_{K \bar K}} }\right) \right] + \mathcal{O}\left(\left|\frac{\lambda_b}{\lambda_s}\right|^2\right)\,.
\end{equation}
The constraints in Fig.~\ref{fig:3a} lead to the bounds in Fig.~\ref{fig:3b}. The CP asymmetry increases with $r_{KK}$ and is inversely proportional to the amount of $U$-spin breaking in $D^0 \to K_{\rm S}^0 K_{\rm S}^0$. Our benchmark scenario, $r_{K K} = 0.5$, corresponds to a CP asymmetry at most at the per-mille level~\cite{Fleischer:2025zhl}.

\section{Conclusions}
We have shown that the measured branching fractions of $D^0\to\pi^-\pi^+$, $D^0\to K^-K^+$ and $D^0\to K_{\rm S}^0K_{\rm S}^0$ can be consistently described within the SM with non-factorisable effects and $U$-spin breaking of order $50\%$. The mode $D^0\to K_{\rm S}^0K_{\rm S}^0$ provides a sensitive probe of these effects. The corresponding direct CP asymmetry is predicted to be at most at the per-mille level, providing a benchmark for future high-precision measurements.

\section*{Acknowledgments}
MLP would like to thank the organisers of Moriond QCD 2026 for the invitation and for the opportunity to present this work. The research presented was supported by the Dutch Research Council (NWO). MLP is supported by the European Union’s Horizon Europe
Research and Innovation Programme under the Marie
Sk{\l}odowska-Curie grant agreement No.~101204923.

\section*{References}

\end{document}